\documentstyle[oldlfont,leqno,twoside,11pt]{article}

\input{psfig}

\newcommand{\COMMENT}[1]{ }

\begin{document}
\cleardoublepage
\pagestyle{myheadings}


\title{Scalable Parallel Numerical Methods\\
       and Software Tools for Material Design\thanks{
This work was supported by ONR contract N00014-93-1-0152,
AFSOR contract F49620-94-1-0286, and ONR contract N00014-91-J-1835.}}
\author{
Eric  J. Bylaska\thanks{Department of Chemistry, University of California,
San Diego.}
\and
Scott R. Kohn\thanks{Department of Computer Science and Engineering,
University of California, San Diego.}
\and
Scott B. Baden\addtocounter{footnote}{-1}\footnote{}
\and
Alan Edelman\thanks{Department of Mathematics, MIT.}
\and
Ryoichi Kawai\thanks{Physics Department, University of Alabama, Birmingham.}
\and
M. Elizabeth G. Ong\thanks{Department of Mathematics,
University of California, San Diego.}
\and
John H. Weare\thanks{To whom correspondence should be addressed.
Department of Chemistry, University of California,
San Diego, 9500 Gilman Drive, La Jolla, CA~92093-0340.
Tel: (619) 534-3286.  Fax: (619) 534-7244.
E-mail: jweare@chem.ucsd.edu.}}
\date{}
\maketitle
\markboth{Bylaska et al.}{Scalable Numerical Methods for Material Design}
\pagenumbering{arabic}


\begin{abstract}
A new method of solution to the local spin density approximation to the
electronic Schr\"{o}dinger equation is presented.  The method is based
on an efficient, parallel, adaptive multigrid eigenvalue solver.
It is shown that adaptivity is both necessary and sufficient to
accurately solve the eigenvalue problem near the singularities at
the atomic centers.  While preliminary, these results suggest that
direct real space methods may provide a much needed method for efficiently
computing the forces in complex materials.
\end{abstract}


\section{Introduction}

To intelligently design materials with specific high performance properties,
it is necessary to have an understanding of the underlying atomic
structure, reactive sites, and other properties of complex candidate compounds.
To achieve the generality and reliability needed to predict
these properties, methods based on the first principles solution
to the electronic Schr\"odinger equation are required.
For systems of typical size, the most reliable and efficient first
principles approach is based on the local density
approximation (LDA) of Kohn and Sham \cite{Kosh65} to the
full many-electron Schr\"odinger equation.
However, current methods of solution scale as $O(N^3)$,
where $N$ is the number of atoms.
For systems of the size commonly encountered in
materials science, such calculations are too large to be practical.

The goal of our program is to develop methods that
can efficiently treat large and complex systems.  To be successful,
we must solve the following computational issues:
\begin{itemize}
{\em
\item The method must be fast to allow simulations requiring
thousands of atomic interaction evaluations.
\item The method must be capable of high accuracy: .02 eV/atom.
\item The method must effectively capture the multiple
length scales  inherent in the problem.
\item The method must scale as $N^2$ or less to allow extension to
larger systems.
}
\end{itemize}

\noindent
To address these goals, we are developing the following techniques
and software tools:
\begin{itemize}
{\em
\item A rapidly converging method for
the non-linear eigenvalue problem arising in the LDA.
\item Adaptive methods for resolving the locality of
electronic wavefunctions with multiple
length scales.
\item A software infrastructure to exploit the high performance parallel
architectures capable of providing the throughput and memory we require.
}
\end{itemize}


\section{The LDA Equations}

In the LDA, the electronic wavefunctions are given by the solutions to the
eigenvalue problem:
\begin{equation}  \label{eq:kohn-sham}
{\cal H} \psi_{i} = \lambda_{i} \psi_{i}
\end{equation}
where the Hamiltonian ${\cal H} $ is given by:
\begin{equation}
\label{eq:kohn-sham2}
{\cal H} = \left ( \frac{- \nabla^{2}}{2m} + V_{ext} + V_{H} + V_{xc} \right )
\end{equation}
$\lambda_{i}$ is an eigenvalue, and the eigenvectors (the wavefunctions
$\psi_i$) satisfy the usual orthonormality constraints of a symmetric operator.
In general, we require the lowest $N$ eigenvalues, where $N$ is the
number of electrons in the system.
Electron-electron interaction is included in the Hartree potential, $V_{H}$,
and the exchange correlation potential, $V_{xc}$.  Both $V_{H}$ and $V_{ex}$
are functions of the charge density $\rho(\vec{x})=\sum_{occ} \psi_i^2$,
where the sum includes only occupied orbitals.
$V_{H}$ is the solution to Poisson's equation in free space with this
charge density.

Since $V_{H}$ and $V_{xc}$ are functionals of the electron density,
Eq.~(\ref{eq:kohn-sham}) must be
solved self-consistently.
That is, an initial density is input and iterations continue until the
input and output densities are the same.  The $V_{ext}$
potential term represents the attractive interactions of the electrons
to the atomic nuclei and is a function of the positions of the atoms.
In our simulations, Eq.~(\ref{eq:kohn-sham})
must be solved many times as the position of the atoms change.

There are several length scales in the solution of
Eq.~(\ref{eq:kohn-sham}).
The overall dimension of the system is determined by the
atomic positions and the associated electron density.
However, each atomic center is
associated with a length related to the effective charge of its
nucleus.
For example, sodium has a small atomic charge and,
therefore, a fairly long length scale ($\approx 2.5$ \AA).  On the other hand,
oxygen has a high effective charge
and a corresponding very short length scale ($\approx .5$ \AA).

The presence of several length scales in Eq.~(\ref{eq:kohn-sham}) poses
significant difficulties for present solution methods, based on the FFT.
Since increases in the overall dimension of the system and the resolution
of the function in real space (because of a short length scale)
both require increases in the size of the basis,
the use of a
planewave basis requires the retention of a very large numbers of basis
functions.
The computational cost of this is somewhat offset by
the high parallelism and efficient vectorization of the
algorithm.
However, because of the steepness of the  atomic potentials,
we have found that on
the order of $10^4$ to $10^6$ Fourier functions may be
required to obtain sufficient accuracy.
Such calculations are extremely CPU intensive.

The eigenvalue equation for a real system is complicated by details
which obscure the essential difficulties of its solution \cite{sung94}.
To develop test problems (see Section~\ref{model}) which retain the
essential singular behavior while removing nonessential details,
we replace $V_H$, $V_{ext}$, and $V_{ex}$ by simple potentials located
at the atomic sites.  The solution to these eigenvalue problems
provide little information as to the convergence properties of the
numerical method with respect to self-consistency or the efficiency
of the solution to the embedded Poisson problem.  However, they enable
us to address the critical issues of multiple length scales and the
singular behavior of the potential.


\section{Parallel Adaptive Solution to the Eigenvalue Problem}

We have developed a parallel adaptive eigenvalue solver (AMG) which
integrates adaptive mesh refinement techniques \cite{bergeroliger:amr}
\cite{bergercolella:shock} with a novel multigrid eigenvalue
algorithm \cite{mccormick:mgsolver}.  To our knowledge, this
is the first time such methods have been combined to solve
materials science problems.

We solve the eigenvalue problem using the multigrid method of
Cai et al.\ \cite{mccormick:mgsolver}.  Given the linear
eigenvalue problem ${\cal H} \psi = \lambda \psi$,
the following efficiently calculates the lowest eigenvalue
and eigenvector:
\begin{center}
\begin{tabbing}
\hspace{40mm}\=\hspace{5mm}\=\hspace{5mm}\=\hspace{5mm}\= \kill
\\
\>let $\psi$ be an initial guess ($\psi \neq 0$)\\
\>repeat\\
\>\>${\cal H}$-normalize $\psi$:  $(\psi, {\cal H} \psi) = 1$\\
\>\>let $\lambda = (\psi, {\cal H} \psi)/(\psi, \psi)$\\
\>\>perform one multigrid V-cycle on $({\cal H} - \lambda I) \psi = 0$\\
\>until $\|({\cal H}-\lambda I) \psi\| < \varepsilon$ (some error tolerance)
\\
\end{tabbing}
\end{center}
Convergence is rapid; for a typical problem, machine precision is
reached within fifteen iterations.  As with most iterative methods,
a good initial guess can significantly speed convergence.  To
calculate eigenvalues other than the lowest, we apply the above
procedure and, after each V-cycle, orthogonalize the candidate
eigenvector against all previously calculated eigenvectors.

\begin{figure*}
\centerline{\psfig{figure=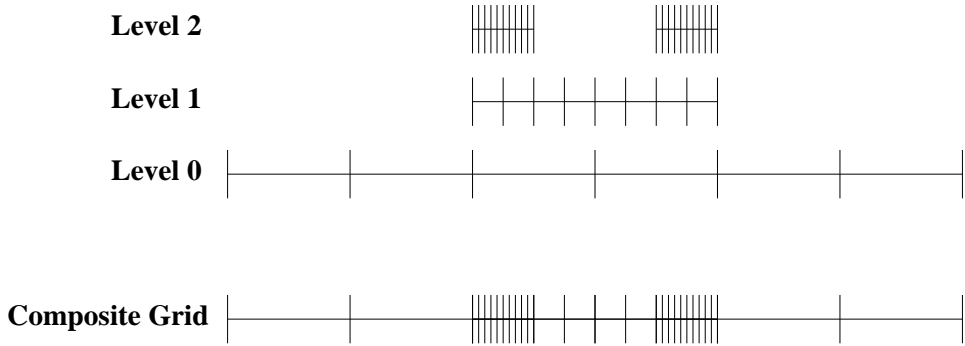,width=5.0in}}
\caption{Wavefunctions are resolved on a composite grid which
represents a non-uniform discretization.  In practice, composite grids
are implemented as a hierarchy of grid levels.}
\label{composite}
\end{figure*}

Because of the multiple length scales present in our problems, we cannot
efficiently represent the eigenvector $\psi$ using a uniform discretization
of space.  Uniform grids cannot adapt in response to local changes; thus,
the grid spacing is dictated by the shortest length scale present in the
{\em entire} problem.  Instead, we represent $\psi$ as a composite grid
(see Figure~\ref{composite}), which enables our solver to {\em locally}
refine the discretization as required by local phenomena.  By exploiting
locality, we expend computational resources (flops and memory) in those
regions of the solution where they are most needed.

A composite grid logically consists of a single grid in which the
discretization is non-uniform.  Such grids are actually represented
using a hierarchy of levels (see Figure~\ref{composite}).  All grids
at the same level have the same mesh spacing, but successive levels
have finer spacing than the ones preceding it, providing a more
accurate representation of the solution.  We locally refine the grid
hierarchy according to an error estimate calculated at run-time.
In general, the location and extent of refinement areas must be
computed by the application, as they cannot be predicted {\em a priori\/}.

We implemented our solver using the LPARX \cite{kohnbaden:LPARX}
parallel programming system, which provides efficient run-time
support for scientific calculations with dynamic, block structured data.
The use of LPARX was essential in facilitating code development; managing
the complicated data structures of a composite grid hierarchy would have
been a daunting task without LPARX, especially on parallel architectures.
LPARX enables us to run the same code on a diversity of high performance
parallel architectures, including the CM-5, Paragon, single processor
workstations, Cray C-90, SP-1, and networks of workstations.  For more
details concerning the implementation and performance, refer to
\cite{kohnbaden:amg} in these proceedings.


\section{Model Problems}
\label{model}

All of the following model problems were solved in 3d;  we did not
attempt to exploit symmetry.  Each AMG solution required approximately
one minute running on an IBM RS/6000 model~590.


\subsection{The Hydrogen Atom}

In this problem, the Hamiltonian has a deceptively simple form with
only a single term:
\begin{equation}
\label{eq:kohn-sham4}
{\cal H} =  -\frac{\nabla^{2}}{2m}-\frac{Z}{r}.
\end{equation}
While the eigenvalue problem corresponding to Eq.~(\ref{eq:kohn-sham4})
can be solved analytically, the singular behavior at $r=0$ can cause
significant problems for numerical methods. In fact, it cannot be
conveniently solved with our present FFT methods.
For example, for the lowest eigenvalue,
our FFT algorithm with $64^3$ mesh points gives the value -0.69 rather
than the correct value of -0.5.
The lowest energy solution in our units is an exponential with the form
$e^{-Zr}$ and energy $E=\frac{Z^2}{2}$.  Note that the severity of the
singularity with increasing $Z$ is reflected in the increasing localization
of the solution around the origin. As $Z$ increases, the density of points
in an adaptive method will increase near the origin.

\begin{figure*}
\centerline{\hbox{
\psfig{figure=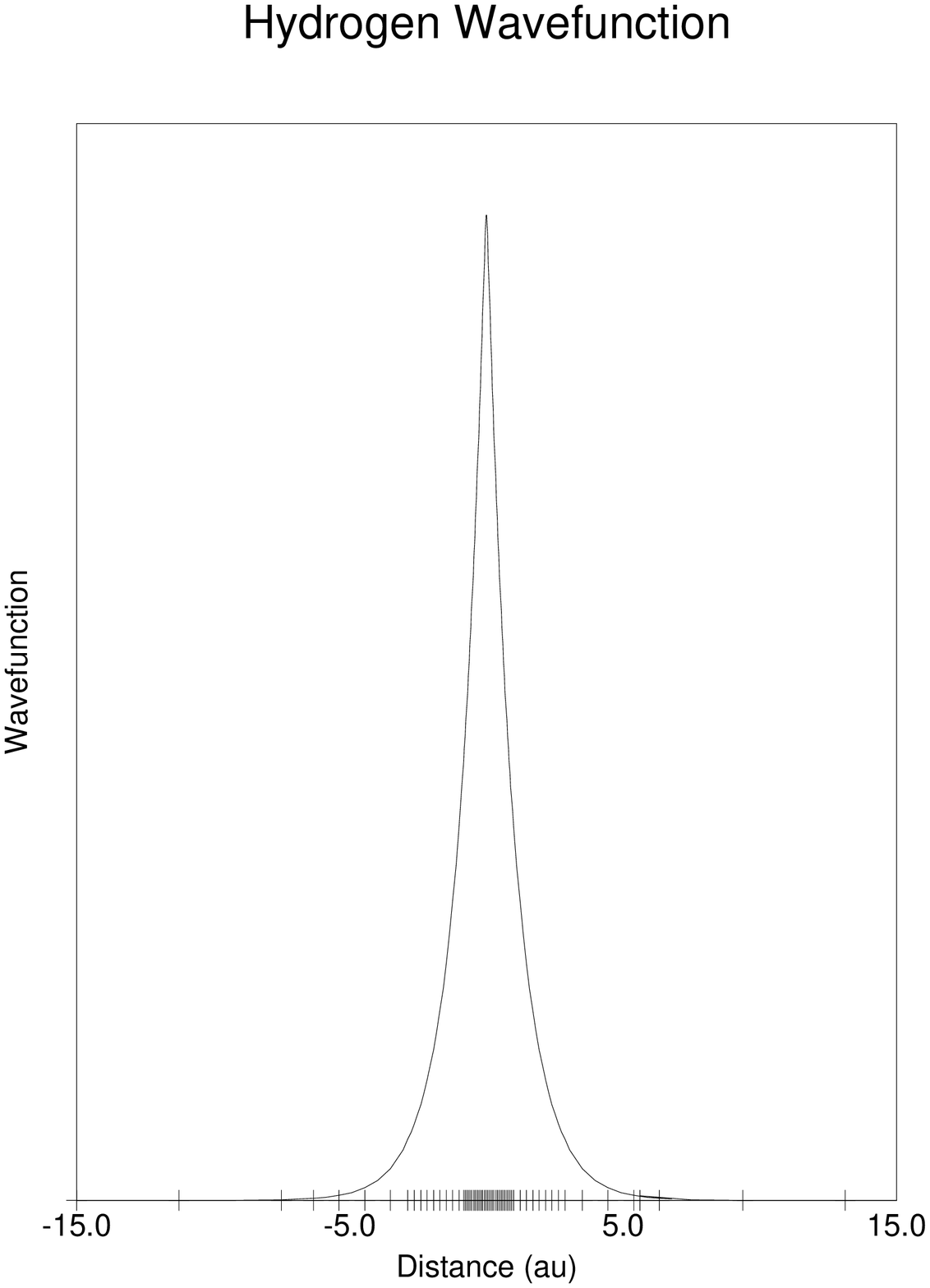,width=2.85in,height=2.6in}
\hspace{0.1in}
\psfig{figure=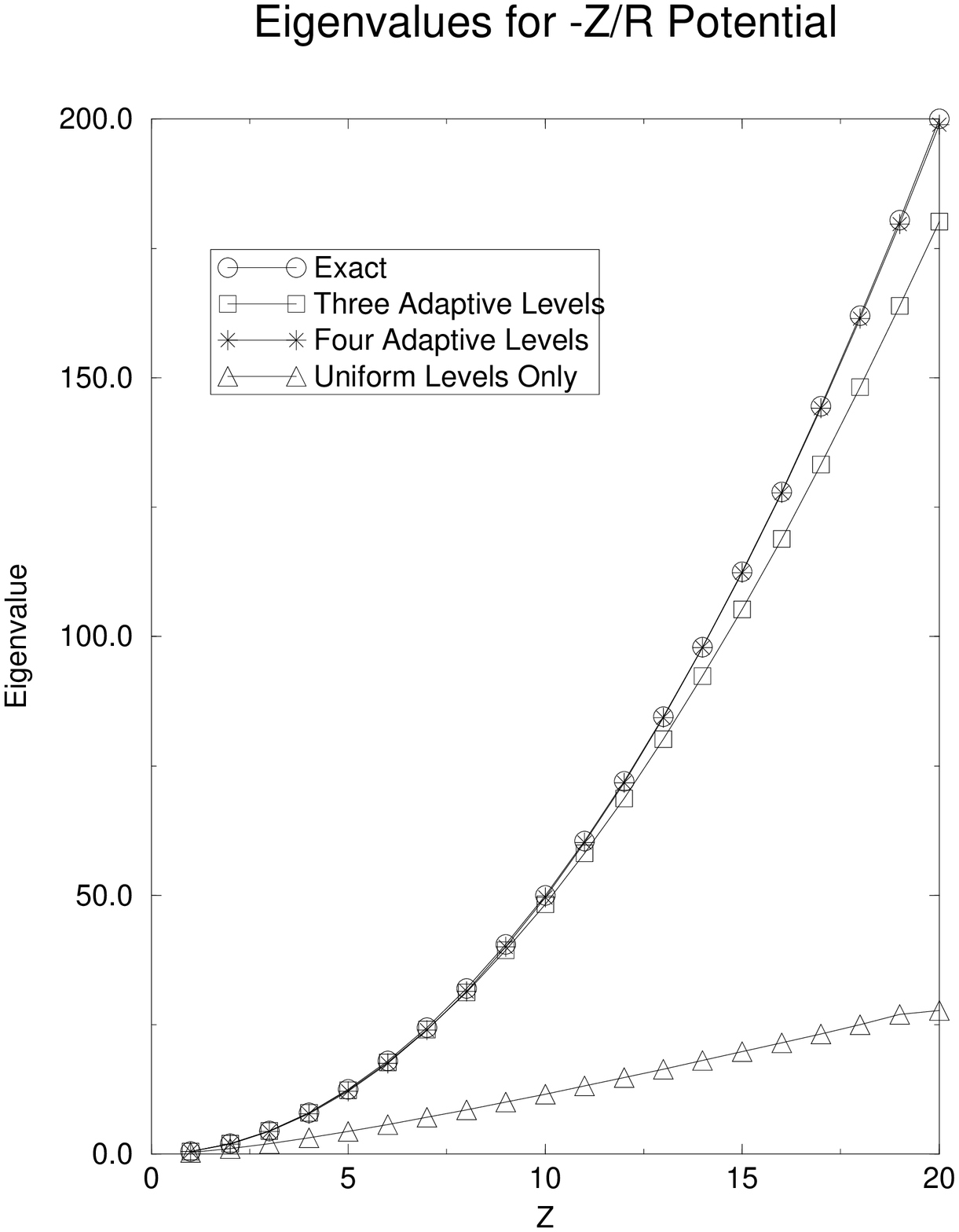,width=2.85in,height=2.6in}}}
\caption{The left graph displays the lowest energy eigenvector
for the hydrogen atom; graph data was extracted from the 3d
volume along the $Z$ axis.  Tick marks on the abscissa represent
mesh points.  The right plot shows the eigenvalues for a
$\frac{-Z}{R}$ potential.}
\label{hydrogen}
\end{figure*}

The $Z=1$ solution corresponds to the hydrogen problem.
It is plotted in Figure~\ref{hydrogen}(a). We note that the AMG solution and
the exact solution (not plotted)
are identical on the scale of the graph.  The cusp at the origin is a
result of the singular nature of the potential at this point.
This behavior is usually difficult to resolve with a numerical
method \cite{arias:wavelet}.

As the singularity strengthens with increasing charge, the lowest energy scales
as $Z^2$.  Figure~\ref{hydrogen}(b) illustrates how this behavior is
reproduced by the AMG solution.  As expected, to obtain the correct scaling,
it is necessary to go to higher levels of adaptivity.  However, because
of increased localization, the total number of points remains roughly
the same.  To illustrate the efficiency of adaptivity, we note that the
resolution at the finest level is equivalent to a uniform
grid with $4096^3$ basis elements, as compared to the fewer than $64^3$
points required by the adaptive algorithm.

\subsection{The $\mbox{H}_2^+$ Molecule}

A problem that is similar to the hydrogen atom problem, but
more commonly used as a test problem for chemical methods, is the
$\mbox{H}_2^+$ molecule.  In this problem, there is only one electron.
However, there are two centers with singularities.  The Hamiltonian is:

\begin{equation}
\label{eq:kohn-sham5}
{\cal H} =  \frac{- \nabla^{2}}{2m}
-\frac{1}{|\vec{r}+\frac{\vec{R_a}}{2}|}
-\frac{1}{|\vec{r}-\frac{\vec{R_a}}{2}|},
\mbox{where $\vec{R_a}$ is the atomic separation.}
\end{equation}

\begin{figure*}
\centerline{\hbox{
\psfig{figure=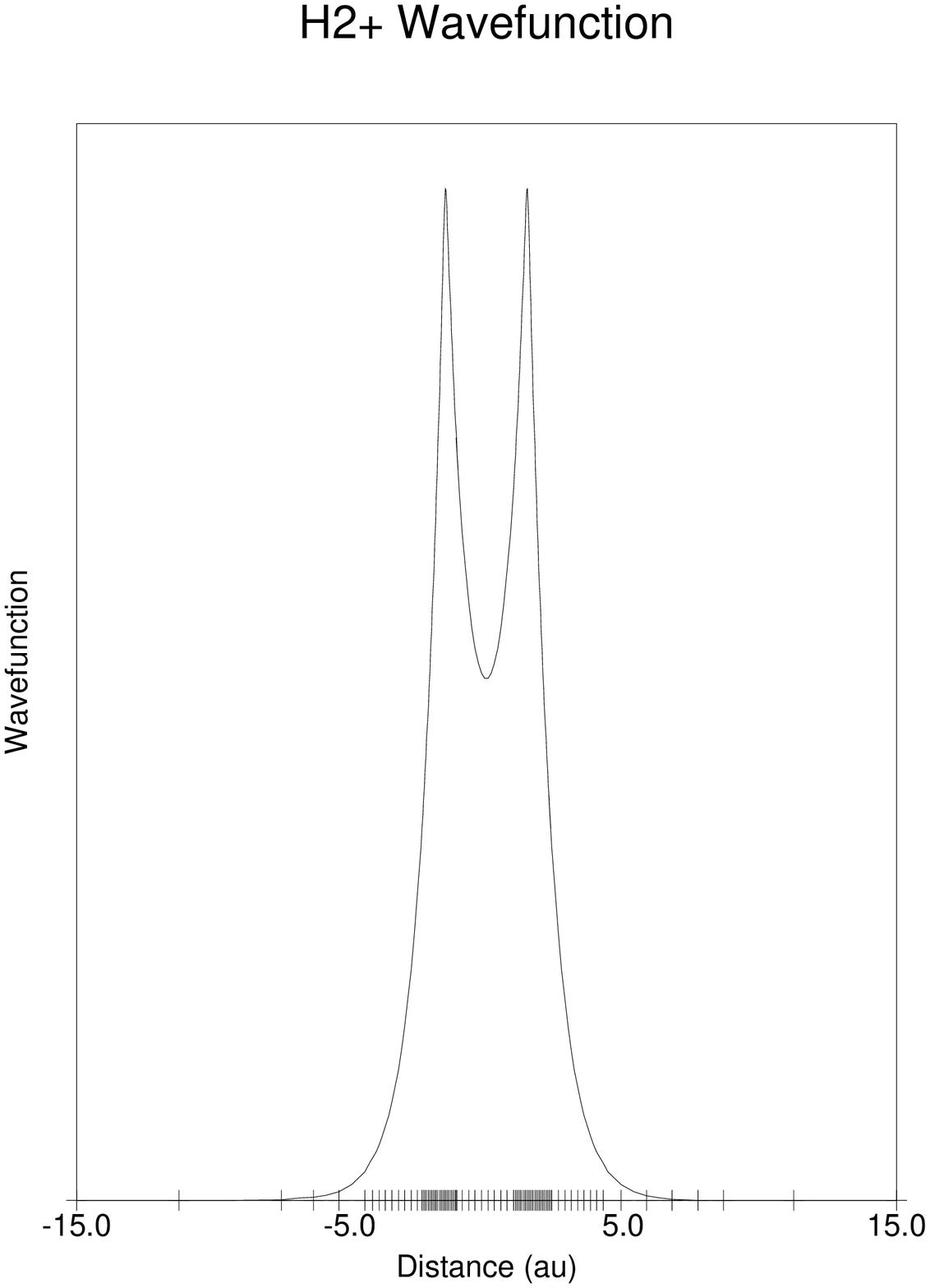,width=2.85in,height=2.6in}
\hspace{0.1in}
\psfig{figure=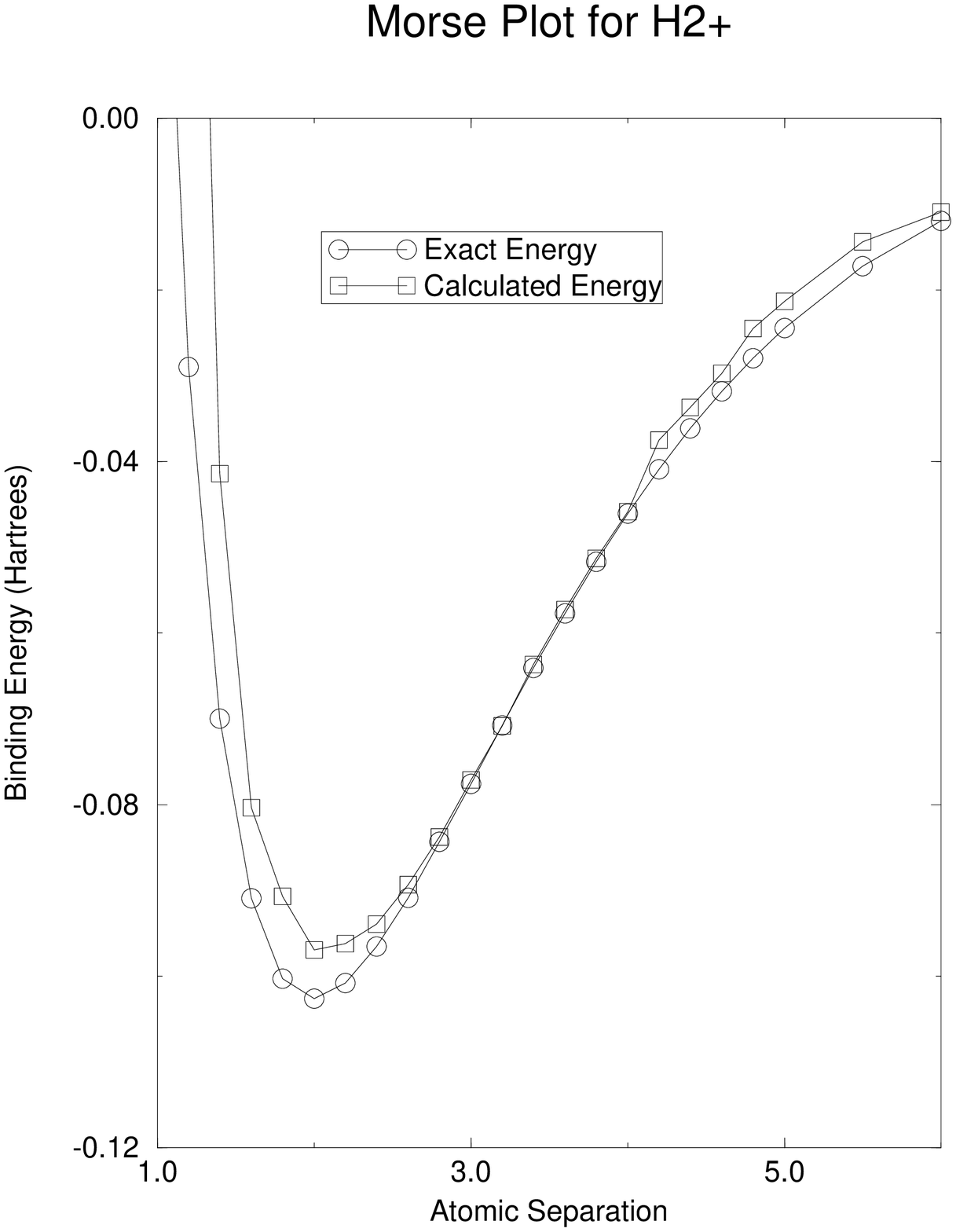,width=2.85in,height=2.6in}}}
\caption{The left graph displays the lowest energy eigenvector for
the hydrogen molecular ion; graph data was extracted from the 3d
volume along the $Z$ axis.  Tick marks on the abscissa represent
mesh points.  The right plot shows binding energy as a function
of atomic separation.}
\label{h2plus}
\end{figure*}

This problem can also be solved analytically \cite{bates:htwoplus}.
Again, it is two stiff for practical solution by FFT.  On the other hand,
the AMG method does quite well as illustrated by the binding energy curve
in Figure~\ref{h2plus}(b).  (Binding energy is
defined as the total energy of the atoms at a specified distance minus the
energy at infinite separation.)  The wave function is plotted in
Figure~\ref{h2plus}(a).  Note the increased density of points in the
vicinity of the nuclei.

\subsection{Adaptive Multigrid vs. FFT}

In this test problem, we soften the singularities in the original potential
by introducing an error function with a variable cut off ($r_{\rm cut}$).
We replace the $\frac{1}{r}$ potentials of Eq.~(~\ref{eq:kohn-sham5})
with the smoothed potentials $erf(\frac{r}{r_{\rm cut}})/r$.
If this potential is sufficiently softened (i.e. $r_{\rm cut}$ is large),
the FFT, the uniform grid, and the AMG methods will all converge
to the same answer. Results are summarized in Table~\ref{rcut}.  The exact
answer for these parameters and $r_{\rm cut}=0$ is~-0.911.  It is clear
that both the uniform grid method and the FFT method lose accuracy
quickly as $r_{\rm cut}$ approaches 0.

\begin{table*}
\caption{A comparison of eigenvalues for the FFT, adaptive multigrid solver,
and a uniform grid solver.  All methods used approximately the same number
of basis elements.  The known solution for $r_{\rm cut}=0$ is~-0.911.}
\label{rcut}
\begin{center}
\begin{tabular}{|l||c|c|c|} \hline
$r_{\rm cut}$ & FFT     & Adaptive & Uniform \\ \hline \hline
0.0           & -1.0946 & -0.9005  & -1.2009 \\
0.1           & -0.9986 & -0.8931  & -1.0353 \\
0.2           & -0.8998 & -0.8734  & -0.9035 \\
0.3           & -0.8664 & -0.8551  & -0.8672 \\
0.4           & -0.8427 & -0.8325  & -0.8430 \\ \hline
\end{tabular}
\end{center}
\end{table*}



\end{document}